Creative Construction Conference 2019, CCC 2019, 29 June - 2 July 2019, Budapest, Hungary

# Assessing Workers' Perceived Risk During Construction Task Using A Wristband-Type Biosensor

Byungjoo Choi[a]*, Gaang Lee[b], Houtan Jebelli[c], SangHyun Lee[b]

[a]Ajou University, 206, World cup-ro, Suwon-si, Gyeonggi-do 16499, South Korea
[b] University of Michigan, 2350 Hayward St., G.G. Brown Building, Ann Arbor, MI 48109, United States
[c] Pennsylvania State University, 104 Engineering Unit A, University Park, PA 16802, United States

## Abstract

The construction industry has demonstrated a high frequency and severity of accidents. Construction accidents are the result of the interaction between unsafe work conditions and workers' unsafe behaviors. Given this relation, perceived risk is determined by an individual's response to a potential work hazard during the work. As such, risk perception is critical to understand workers' unsafe behaviors. Established methods of assessing workers' perceived risk have mainly relied on surveys and interviews. However, these post-hoc methods, which are limited to monitoring dynamic changes in risk perception and conducting surveys at a construction site, may prove cumbersome to workers. Additionally, these methods frequently suffer from self-reported bias. To overcome the limitations of previous subjective measures, this study aims to develop a framework for the objective and continuous prediction of construction workers' perceived risk using physiological signals [e.g., electrodermal activity (EDA)] acquired from workers' wristband-type biosensors. To achieve this objective, physiological signals were collected from eight construction workers while they performed regular tasks in the field. Various filtering methods were applied to exclude noises recorded in the signal and to extract various features of the signals as workers experienced different risk levels. Then, a supervised machine-learning model was trained to explore the applicability of the collected physiological signals for the prediction of risk perception. The results showed that features based on EDA data collected from wristbands are feasible and useful to the process of continuously monitoring workers' perceived risk during ongoing work. This study contributes to an in-depth understanding of construction workers' perceived risk by developing a noninvasive means of continuously monitoring workers' perceived risk.

## 1. Introduction

Despite continuous efforts to reduce construction accidents, safety in construction still lags behind other comparable industries in most countries. For example, in the United States, there were 971 fatal occupational injuries in the construction industry in 2016 [1]. While the construction industry hired 4.3% of employees, it accounted for more than 18% of total work-related deaths in 2017 [1]. Also, there were more than 1,981,000 nonfatal occupational injuries and illnesses in the construction industry, with an incident rate that is 10% higher than the national average incident rate in 2017 [2]. In Korea, the construction industry reported 579 fatal occupational injuries with a fatal incident rate 3.2 times higher than the national average [3]. Furthermore, the poor performance of safety in the construction industry has been a significant issue in other areas including Europe [4], Asia [5], and Australia [6] as well.

Accident investigations have revealed that construction accidents are caused by the interaction between unsafe conditions (i.e., physical environment that contains potential hazard) and unsafe behaviors (i.e., actions that deviate from the safety procedure) [7]. When construction workers are exposed to an unsafe condition, they determine their response to the risk (i.e., safe or unsafe behavior) by comparing the perceived risk with their internal acceptable risk [8]. As such, risk perception is the process in which unsafe conditions interact with workers' safety decision-making processes, which ultimately results in safe or unsafe behaviors. In this regard, there have been continuous efforts to assess workers' perceived risk in construction sites. Such efforts have mainly relied on workers' subjective answers to a set of fixed questions (i.e., survey) [9]. However, these methods are not capable of capturing changes in workers' perceived risk over time because it is almost impossible to conduct the survey at every change in perceived risk. Considering the dynamic nature of construction sites, capturing changes in perceived risk is critical to understand workers' perceived risk. Also, since answers to the questionnaire rely on imprecise memory and the reconstruction of feeling in the past, these methods may be subject to biases [10]. Finally, participating in the survey could interfere with workers' ongoing work because subjects must devote time and energy to answering the questions. For these reasons, there has been increased attention to continuous, objective, and non-intrusive methods to assess workers' perceived risk in construction sites [11].

Recent advancements in wearable biosensors, which can continuously collect an individual's physiological signals [e.g., Photoplethysmography (PPG), Electrodermal activity (EDA), Skin Temperature (ST)], has immense potential to overcome the limitations of the survey-based method. Since physiological signals are affected by the sympathetic nervous system which can be aroused by perceived risk, changes in physiological signals could be indicative of an individual's perceived risk [12]. As such, physiological signals collected from wearable biosensors can be used for continuous, objective, and non-intrusive assessment of construction workers' perceived risk during their ongoing work. Despite such potential of physiological signals acquired from wearable biosensors, their capability to assess workers' perceived risk remains questionable.

To bridge the knowledge gaps, this study aims to develop a framework that recognizes workers' perceived risk based on signal processing and machine learning techniques using physiological signals collected from wearable biosensors during workers' ongoing work. Among diverse physiological signals, this study focuses on EDA, which refers to electro properties of the skin, because EDA is the only physiological signal related to the sympathetic nervous system that is not contaminated by the parasympathetic nervous system. In this study, the authors extracted relevant time and frequency domain features in EDA signals collected from construction workers' wearable biosensors while working on the real construction sites. Then, several supervised learning algorithms were applied to select the best classifiers to recognize workers' perceived risk while working on activities with different risk levels.

## 2. Perceived Risk Recognition

Figure 1 shows an overview of an EDA-based perceived risk recognition procedure developed in this study. The first step is to collect EDA signals from wearable biosensors. At the same time, workers' activities are recorded to label risk levels of the activities. The second step is to remove the artifacts included in the signal and to decompose the EDA signal into tonic [i.e., Electrodermal Level (EDL)] and phasic component [i.e., Electrodermal Response (EDR)). After removing the artifacts and decompose the signal, time EDL and EDR features in time and frequency domains are extracted. Then, the authors applied several supervised learning algorithms [i.e., Decision Tree (DT), Logistic Regression (LR), Gaussian Support Vector Machine (GSVM), K- Nearest Neighbor (KNN), Subspace KNN (SKNN), and Banging Tree (BT)] to identify the best classifier to recognize workers' perceived risk during their ongoing work. Details of each step in the developed framework will be demonstrated in the following sub-sections.

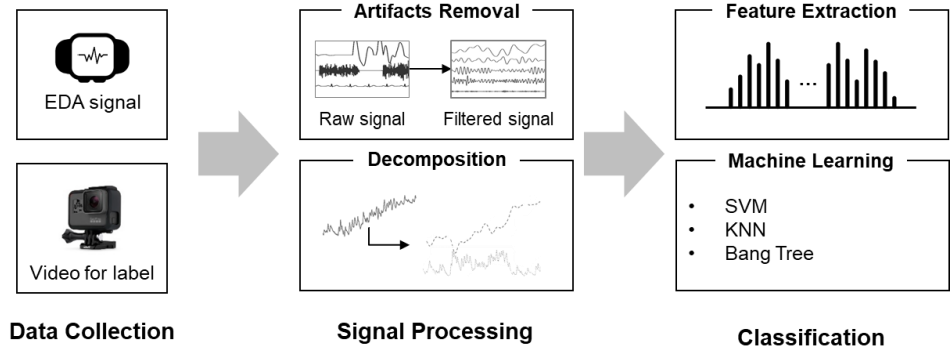

Figure 1. Perceived Risk Recognition Framework

## 2.1. Data Preprocessing – Artifacts Removal and EDA Decomposition

EDA signals collected by a wearable sensor from construction sites contain a large number of artifacts. An artifact is defined as any variation in the recorded signals that does not originate from the signal source of interest [13]. Artifacts removal is critical to recognize workers' perceived risk because artifacts distort the original signals and skew the analysis [13]. A second-order high-pass filter (cut-off frequency $f_c = 0.05$ Hz) is applied to alleviate the artifacts low-frequency noises (e.g., noises introduced from variations in temperature and humidity). Also, moving average filter is applied to mitigate high-frequency noises caused by excessive movements and electrode popping, and adjustment of sensors[14].

As aforementioned, EDA signals can be categorized into tonic (i.e., EDL) and phasic (i.e., EDR) components. EDL represents slow changes in the signal which shows baseline drift. EDR refers to short-term changes in EDA signals that reflect an immediate response to external stimuli. To decompose EDA into EDL and EDR, the authors apply the convex optimization method [15]. This method takes apart EDL and EDR from Noisy EDA based on prior knowledge about patterns of EDL and EDR.

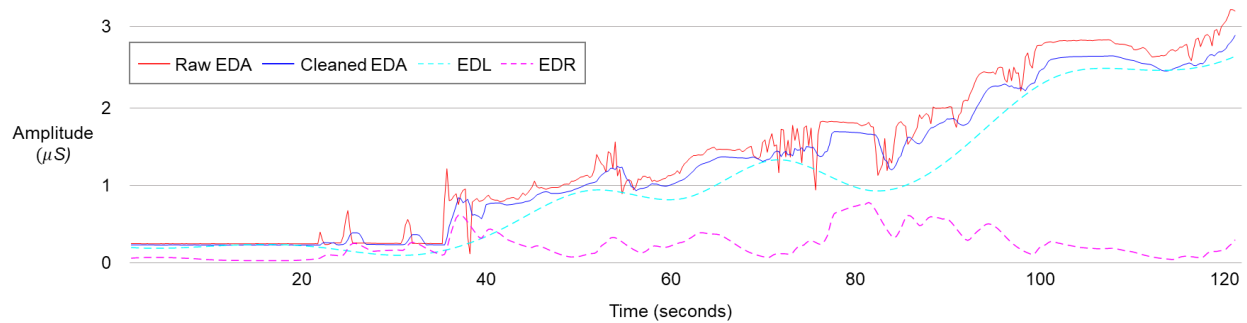

Figure 2. Data Pre-processing: Artifacts Removal and Decomposition

## 2.2. Feature Extraction

After removing artifacts and decomposing the EDA signals, the authors extracted features that have been used in the literature to recognize the EDA signal patterns reflecting individuals' physiological response to the external stimuli. To extract features, time-series decomposed signals (i.e., EDL and EDR) are segmented using 10 seconds blocks with 9 seconds overlapping (i.e., 1-second moving). The window size is set to 10 seconds considering time lags between the external stimuli and physiological responses. From the blocks of 10 seconds, the authors calculated eight time-domain features and three frequency-domain features. Since EDR reflects immediate responses to external stimuli, features are mainly extracted from EDR but several features from EDL are included because EDL shows general changes in EDA signals that can vary across individuals. Table 1 shows the extracted features in this study.

Table 1. Selected Features in time and frequency domain

| Domain | Features |
|---|---|
| Time domain | Standard deviation of EDR, Median of EDR, Integral of EDR, Normalized average power of EDR, Normalized root mean square of EDR, Mean of EDL, Standard deviation of EDL, Median of EDL |
| Frequency domain | Spectral power of band 1 (0.1Hz ~ 0.2Hz) of EDR, Spectral power of band 2 (0.2Hz ~ 0.3Hz) of EDR, Spectral power of band 3 (0.3Hz ~ 0.4Hz) of EDR |

*2.3. Classification*

With the extracted features, several classification algorithms are tested to select the best classification algorithm. First, DT, LR, GSVM, and KNN, which have been used in previous studies to detect sympathetic arousal, are tested. In addition, the authors tested two ensemble classification algorithms (i.e., BT, and SKNN) that can facilitate a more accurate performance than their non-ensemble counterpart (i.e., DT, and KNN). To select the best classification algorithm, the tested algorithms are compared in terms of test accuracy. Specifically, data is first undersampled to make the two classes (i.e., high risk and low risk) balanced. When the EDA samples are imbalanced, which means that the number of samples of one class is much more than that of other class, the resultant classifier tends to be inaccurate for predicting the minority class. To avoid such a problem, the majority class is randomly undersampled to make the two classes have the same number of samples. Then, the undersampled data is randomly divided into 80% training session and 20% testing session without overlap. Using the training session, classifiers are trained, and then the trained classifiers' accuracy is calculated by counting how many samples in the testing session are correctly classified by the trained classifiers. This undersample-train-test procedure is repeated 20 times and accuracy is averaged to ensure that the resultant accuracy shows the general performance, not the performance in a randomly generated subset.

## 3. Field Data Collection

To examine the feasibility of the developed framework in recognizing workers' perceived risk, a field data collection is conducted. EDA signals are collected from eight workers during their ongoing work. All subjects are male and reported no physical and mental disorders that can affect their daily activities. Before starting the data collection, all subjects are informed about the data collection procedure and asked to provide their information including age, gender, height, weight, work experience, and trade. Table 2 summarize demographic information of the subjects.

Table 2. Demographic information of the subjects

| Domain | Age (years) | Height (cm) | Weight (kg) | Work Experience (years) |
|---|---|---|---|---|
| Mean | 32.37 | 181.5 | 88.88 | 10.25 |
| Standard Deviation | 8.57 | 7.14 | 12.92 | 6.72 |

After providing the demographic information, all subjects are asked to wear a wristband-type biosensor to collect EDA data during their ongoing work. After wearing the sensor, subjects are asked to perform their daily work. At the same time, subjects' activities are video recorded using an action camera to label the activities at different risk levels (i.e., low-risk and high-risk). Each worker participates in the data collection during half of their working hours. The data collection procedures are approved by the IRB (Internal Review Board) of the University of Michigan. After the data collection, two research team members, who have expertise in risk related to construction activities, separately classify all subjects' activities into low-risk activities and high-risk activities using the recorded video. Figure 2 shows the wristband used in the data collection and examples of low and high-risk labeled activities in the data collection. In the case of inconsistency in the results of the label between two members, the data were excluded in the analysis to improve the reliability of the analysis. The labeled data are used to learn the machine learning model and evaluate the performance of the classifiers.

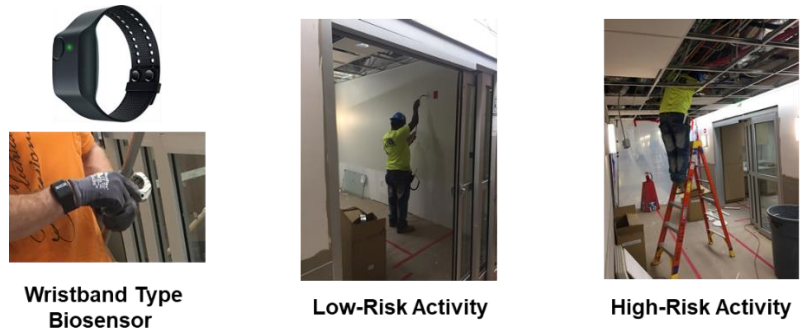

**Wristband Type Biosensor**          **Low-Risk Activity**          **High-Risk Activity**

Figure 3. Wristband and Examples of Data Labelling

## 4. Results and Discussion

As aforementioned, six classification algorithms are compared. The accuracy indicates the algorithms' overall performance to correctly classify the EDA samples. Table 3 shows the test accuracy of the tested algorithms. As a result, KNN showed the best accuracy, 76.9%. Since KNN, which does not learn a decision boundary based on parametric models, outperformed other parametric algorithms such as LR and GSVM, this result indicates that the data is hard to classify by parametric models in the given dimension (i.e., 11 EDA features). When additional features that can help to separate the two classes are added, the parametric algorithms might be able to outperform KNN.

Table 3. Classification accuracies of each tested algorithm

| Algorithm | DT | LR | GSVM | KNN | BT | SKNN |
|---|---|---|---|---|---|---|
| Accuracy | 69.1% | 61.7% | 73.2% | 76.9% | 76.4% | 71.8% |

In addition to the accuracy, the authors computed precision and recall to investigate whether the classifiers are biased to accurately work for only one class. Precision shows the performance of the classifier to exclude actual low-risk samples from high-risk class, while recall shows the performance to include actual high-risk samples in high-risk class. Since the data was first undersampled to make the two classes have a sample number of samples, the prevision and recall of the tested algorithms were quite close to accuracy as shown in Table 4. However, recall is generally higher than precision throughout all the tested algorithms. This may be explained by the limitation of the labelling method. Since research team members labelled the level of risk based on the research team members' subjective evaluations based on the recorded video, if the subjects perceived high risk by other stimuli that were hard to be captured by video (e.g., risk factors that are visually perceived by the subjects, but out of video angle), actually high risk samples can be mislabelled as low risk.

Table 4. Confusion Matrices

| DT | | True Class | | Precision | Recall |
|---|---|---|---|---|---|
| | | High Risk | Low Risk | | |
| Predicted Class | High-Risk | 41,223 | 19,900 | 67.4% | 73.7% |
| | Low-Risk | 14,677 | 36,000 | | |

| KNN | | True Class | | Precision | Recall |
|---|---|---|---|---|---|
| | | High Risk | Low Risk | | |
| Predicted Class | High-Risk | 44,664 | 14,580 | 75.4% | 79.9% |
| | Low-Risk | 11,236 | 41,320 | | |

| LR | | True Class | | Precision | Recall |
|---|---|---|---|---|---|
| | | High Risk | Low Risk | | |
| Predicted Class | High-Risk | 37,242 | 24,211 | 60.6% | 66.6% |
| | Low-Risk | 18,658 | 31,689 | | |

| BT | | True Class | | Precision | Recall |
|---|---|---|---|---|---|
| | | High Risk | Low Risk | | |
| Predicted Class | High-Risk | 46,017 | 16,552 | 73.5% | 82.3% |
| | Low-Risk | 9,883 | 39,348 | | |

| GSVM | | True Class | | Precision | Recall |
|---|---|---|---|---|---|
| | | High Risk | Low Risk | | |
| Predicted Class | High-Risk | 43,471 | 17,557 | 71.2% | 77.8% |
| | Low-Risk | 12,429 | 38,343 | | |

| SKNN | | True Class | | Precision | Recall |
|---|---|---|---|---|---|
| | | High Risk | Low Risk | | |
| Predicted Class | High-Risk | 41,106 | 16,782 | 71.0% | 73.5% |
| | Low-Risk | 14,794 | 39,118 | | |

## 5. Conclusion

This study examines the feasibility of using physiological signals (EDA) acquired from a wristband type biosensor to recognize construction workers' perceived risk at construction site by applying signal processing methods and developing a machine learning model. To select the best classifier, this study evaluates the performance of several supervised learning algorithm (i.e., DT, LR, GSVM, KNN, SKNN, and BT) in recognizing workers' perceived risk during their ongoing work. KNN shows the best performance with an accuracy of 76.9%. The results show that features based on EDA signals have capability to recognize construction workers' perceived risk while working under different risk levels (i.e., low and high-risk activities) at real construction sites. The results contribute to an in-depth understanding of construction workers' perceived risk by developing a non-invasive means of continuously monitoring workers' perceived risk. The proposed framework can contribute to improving construction safety continuously monitoring of workers' perceived risk during their ongoing work. Further research should be conducted to optimize the performance of the framework by comparing different window sizes and including features extracted from other physiological signals such as PPG and ST.